\documentclass[prd,onecolumn,superscriptaddress,preprintnumbers,nofootinbib,APS]{revtex4}

\usepackage{graphicx}
\usepackage{epsf}
\usepackage{amsmath}
\usepackage{epstopdf}
\usepackage{bm}
\usepackage{color}
\usepackage{tabularx}
\usepackage{enumitem}
\usepackage{float}
\usepackage{array,booktabs}
\usepackage{footnote}
\usepackage{threeparttable}
\usepackage{graphicx}
\usepackage{hyperref}
\usepackage{amssymb,epsf}
\usepackage{latexsym}
\usepackage{epstopdf}
\usepackage{epsfig}

\begin{document}
\title{Absence of isolated critical points with nonstandard critical exponents\\
in the four-dimensional regularization of Lovelock gravity}

\author{Ali Dehghani\footnote{email adress: ali.dehghani.phys@gmail.com}} \affiliation{Department of Physics and Biruni Observatory, College of Sciences, Shiraz University, Shiraz 71454, Iran}
 
\author{Mohammad Reza Setare\footnote{email address: rezakord@ipm.ir}}
\affiliation{Department of Physics, Campus of Bijar, University of Kurdistan, Bijar, Iran}

\begin{abstract}
Hyperbolic vacuum black holes in Lovelock gravity theories of odd order $N$, in which $N$ denotes the order of higher-curvature corrections, are known to have the so-called isolated critical points with nonstandard critical exponents (as $\alpha = 0$, $\beta = 1$, $\gamma = N-1$, and $\delta = N$), different from those of mean-field critical exponents (with $\alpha = 0$, $\beta = 1/2$, $\gamma = 1$, and $\delta = 3$). Motivated by this important observation, here, we explore the consequences of taking the $D \to 4$ limit of Lovelock gravity and the possibility of finding nonstandard critical exponents associated with isolated critical points in four-dimensions by use of the four-dimensional regularization technique, proposed recently by Glavan and Lin \cite{Glavan2020}. To do so, we first present $\text{AdS}_4$ Einstein-Lovelock black holes with fine-tuned Lovelock couplings in the regularized theory, which is needed for our purpose. Next, it is shown that the regularized $4D$ Einstein-Lovelock gravity theories of odd order $N > 3$ do not possess any physical isolated critical point, unlike the conventional Lovelock gravity. In fact, the critical (inflection) points of the equation of state always occur for the branch of black holes with negative entropy. The situation is quite different for the case of the regularized $4D$ Einstein-Lovelock gravity with cubic curvature corrections ($N=3$). In this case ($N=3$), although the entropy is non-negative and the equation of state of hyperbolic vacuum black holes has a nonstandard Taylor expansion about its inflection point, but there is no criticality associated with this special point. At the inflection point, the physical properties of the black hole system change drastically, e.g., both the mass and entropy of the black hole vanishes, meaning that there do not exist degrees of freedom in order for a phase transition to occur. These results are in strong contrast to those findings in Lovelock gravity.
\end{abstract}

\maketitle

 \section{Introduction} \label{sect1:intro}

Einstein's general relativity as well as its generalizations in the presence of a cosmological constant (positive $\Lambda> 0$ or negative $\Lambda < 0$) or in the absence of a cosmological constant ($\Lambda = 0$) predict the so-called black hole spacetimes \cite{Carroll,Zee}. The prediction of black holes is one of the fascinating phenomena in theoretical physics, which has now been confirmed by empirical evidence \cite{Genzel1996,LIGO2016,LIGO2017}. It is believed that these mysterious objects can provide a basis for studying the quantum effects in gravity and perhaps a path to quantum gravity \cite{QGMukhanov,QGBurgess,QGHartman,QGRovelli}. In this regard, we can refer to the study of scalar quantum fields in a Schwarzschild black hole background, which led to the discovery of Hawking's thermal radiation from black holes \cite{Hawking1975}. This discovery paved the way for a thermodynamical definition of entropy by evaluating the semi-classical partition function \cite{GibbonsHawking1977}, which exactly matches the Bekenstein entropy \cite{Bekenstein1972}. Having this essential ingredient, one can formulate the four laws of black hole mechanics which correspond to the four laws of thermodynamics \cite{Bardeen1973}, specially the first law for a rotating charged (Kerr-Newman) black hole may be written as
 \begin{equation}
 dM = TdS + \Phi dQ + \Omega dJ,
 \end{equation}
 where $M$ is the ADM mass of the black hole, $T$ is the temperature, $S$ is the entropy, $\Phi$ is the potential, $Q$ is the charge, $\Omega$ is the angular velocity, and $J$ is the angular momentum. However, the major difference in comparison with everyday thermodynamic systems is that there is no pressure-volume ($P$-$V$) term in the first law of black hole thermodynamics. Subsequently, in the last decade, novel innovation in the subject of black hole thermodynamics have occurred due to the inclusion of the pressure-volume term into the first law of black hole thermodynamics \cite{Kastor2009,Kastor2010,Dolan2011a,Dolan2011b,KubiznakMann2012}, which is widely called the extended phase space thermodynamics (see \cite{CQG2017Review} for a nice review). Treating the cosmological constant as thermodynamic pressure, i.e., $\Lambda=-8 \pi P$, is the primary assumption of the extended phase space thermodynamics of AdS black holes \cite{Kastor2009,Kastor2010}. This leads to consistency of both the first law of thermodynamics and the corresponding Smarr relation, meaning that every dimensionful parameter in the theory has a thermodynamic interpretation and such parameters should be treated as thermodynamic variables \cite{Kastor2010,CQG2017Review}. In this modern context, the extended first law is derived as \cite{Kastor2009,CQG2017Review}
  \begin{equation}
 dM = TdS + VdP+\Phi dQ + \Omega dJ,
 \end{equation}
indicating that the ADM mass has to be identified with the chemical enthalpy of the system. This modern viewpoint has opened up a whole new world of black hole thermodynamics, including new concepts such as critical phenomena and the associated thermodynamic phase transitions, reverse isoperimetric inequality, holographic heat engines, Joule-Thomson expansion, black hole molecule, etc. \cite{KubiznakMann2012,CQG2017Review,Mann2012Altamirano,Mann2014TriplePoint,HennigarMann2017PRL,Astefanesei2019,PRD2020,RevIso2011,HE2014,HE2017a,HE2019b,HE2021,JT2017,JT2018a,JT2018b,JT2021,Wei2015,Karch2015}. \vspace{2mm}

At present, a great deal of research has been devoted to exploring $P-v$ criticality and the corresponding thermodynamic phase transitions of various charged-AdS black hole spacetimes (see \cite{KubiznakMann2012,CQG2017Review,Mann2012Altamirano,Mann2014TriplePoint,HennigarMann2017PRL,Astefanesei2019,PRD2020} and references therein). The first step in analyzing the criticality of a thermodynamic system is to find the corresponding critical points. The physics of critical points is well described by critical exponents which determine the behavior of thermodynamic quantities during continuous phase transitions \cite{HuangBook}. It is believed that these exponents do not depend on the microscopic details of a physical system and they are highly affected by only a very few quantities, such as spatial dimensions, symmetries, and the number of components of the order parameter \cite{HuangBook}. This is an important example of universality in physics. \vspace{2mm}

Remarkably, the same as critical exponents in everyday thermodynamical systems, a wealth of evidence indicate that phase transitions associated with second-order critical points arising in different black hole spacetimes often possess the same set of mean field theory critical exponents as $\alpha = 0$, $\beta = 1/2$, $\gamma = 1$, and $\delta = 3$ \cite{KubiznakMann2012,CQG2017Review,Mann2012Altamirano,HennigarMann2017PRL,CQG2020,DH2021,Hennigar2017JHEP,Zou2014,Galaxies2014}. However, there exist some examples of physical systems with critical exponents deviating from those of the mean field theory, e.g., in anisotropic systems (in $D \le 5$ dimensions) as well as lower-dimensional ($2D$ and $3D$) Ising models \cite{NonStaCritExp1996,NonStaCritExp1974,NonStaCritExp1992,NonStaCritExp2004,NonStaCritExp2008}. In black hole physics, the only exception so far found is for hyperbolic vacuum black holes in Lovelock gravity theories \cite{Lovelock1971,Lovelock1972} (as well as in its generalizations) of odd order $N$ with nonstandard critical exponents as $\alpha = 0$, $\beta = 1$, $\gamma = N-1$ $\delta = N$ associated with the so-called isolated critical points \cite{ICP2014,Frassino2014Mann,Brenna2015,HennigarTjoa2017,Dykaar2017JHEP,Mir2019a,Mir2019b}. In these theories, $N$ is the maximum order of higher-order curvature terms and such terms appear in higher dimensions ($D \ge 5$), meaning that there exist no upper critical dimension (which often appears in field theory) above which the critical exponents of the theory become the same as those in mean field theory. \vspace{2mm}

The question is whether these results can be achieved from a four-dimensional black hole spacetime or not. It would be remarkable to find a gravitational theory that reproduces these results in four dimensions. Here, we investigate this question within the framework of the so-called regularized $4D$ Einstein-Lovelock gravity. Recently, a simple dimensional regularization of Lovelock gravity has been discussed in a number of research papers by which the effects of ghost-free, higher-order curvatures can be extended to four-dimensional spacetimes \cite{Tomozawa2011,Zerbini2013,Glavan2020,Casalino2021,Zhidenko2020a,Gao2021}. As a result of this dimensional reduction, it is expected that some features of Lovelock gravity may be observed again, or may be lost, or may undergo fundamental changes in four dimensions. Since all the known examples of nonstandard critical exponents are relied on the correction terms arisen from Lovelock Lagrangian of odd order in curvature \cite{ICP2014,Frassino2014Mann,HennigarTjoa2017,Dykaar2017JHEP,Mir2019a,Mir2019b,Brenna2015} and naturally appear in spacetime dimensions more then four, it becomes of interest to investigate the consequence of taking the $D \to 4$ limit of such Lovelock gravity theories on this feature. \vspace{2mm}

It should be emphasized that the regularized $4D$ theory we are dealing with in the present paper is a method developed by Glavan and Lin \cite{Glavan2020,Casalino2021,Zhidenko2020a} based on rescaling the Lovelock coupling constants at the level of field equations (see eq. (\ref{rescaling}) in Sect. \ref{sect2:Regularized Lovelock}) and then taking the $D \to 4$ limit of the resultant field equations which, for example up to the leading order of corrections, leads to a $D \to 4$ limit of solutions of the $D$-dimensional Gauss-Bonnet gravity \cite{Glavan2020}. While the resultant novel theory enjoys a number of remarkable properties such as (presumably) bypassing the Lovelock's theorem, avoiding Ostrogradski instability, preserving the two standard polarizations of the massless graviton as well as successfully mimicking some consequences of Lovelock gravity in four-dimensions, but it sounds ambitious in some aspects \cite{Tekin2020,Mann2020EGB4D,Mahapatra,Ai2020,Hinterbichler2020,Kobayashi2020,LuPang2020,Mukohyama2020} which have caused debates about the correct $4D$ regularization procedure of Lovelock theory (for a nice review see Ref. \cite{EGB4Dreview}). The authors of Ref. \cite{Tekin2020} by splitting the Gauss-Bonnet tensor disclosed that Einstein-Gauss-Bonnet gravity in four dimensions does not have an intrinsically $4D$ description in terms of a covariantly-conserved rank-2 tensor, concluding that there is no pure $4D$ Gauss-Bonnet gravity without introducing any extra fundamental degrees of freedom. Although the Glavan and Lin's theory is presented at the level of the field equations, but analysis at the level of action also revealed that unphysical divergences appears which make the theory ill-defined unless some surface terms and counterterms are added to the action \cite{Mahapatra}. In Ref. \cite{Mann2020EGB4D}, it has been shown that a naive $D \to 4$ limit of $D$-dimensional Gauss-Bonnet gravity upon rescaling the Gauss-Bonnet coupling constant (as $\alpha \to \frac{\alpha}{D-4}$) is not unique and a well-defined $D \to 4$ limit of conventional Gauss-Bonnet gravity can be achieved by implementing the so-called conformal regularization \cite{Mann1993}. More importantly, the scalar-tensor reformulation of regularized $4D$ Lovelock gravity has been presented in Refs. \cite{Kobayashi2020,LuPang2020} by means of Kaluza-Klein reduction, in agreement with the results of Ref. \cite{Hinterbichler2020}, in which the authors have proved that taking the $D \to 4$ limit of $D$-dimensional Gauss-Bonnet amplitudes leads to the amplitudes of a scalar-tensor theory (i.e., a subclass of Horndeski theory \cite{Horndeski1974}). Another method is the temporal diffeomorphism breaking regularization in which the theory is unique up to a choice of a constraint that stems from a temporal gauge condition \cite{Mukohyama2020}. The aforementioned alternative regularized $4D$ theories indicate that additional degrees of freedom are necessary to have well-defined field equations and, interestingly, all of them respect Lovelock's theorem. In Sect. (\ref{sect4:conclusion}), we comment on whether the results of this paper may hold for the alternative $4D$ regularizations of Lovelock gravity or not. \vspace{2mm}

The organization of the paper is as follows. In Sec. \ref{sect2:Regularized Lovelock}, we present the $\text{(A)dS}_4$ Einstein-Lovelock black hole solutions with finely tuned Lovelock couplings and then the thermodynamics of $\text{AdS}_4$ black hole solutions in the general (not only the finely tuned) case is reviewed. Next, in Sec. \ref{sect3:Absence}, we study the equation of state of $\text{AdS}_4$ Einstein-Lovelock black holes with hyperbolic symmetry in order to investigate the possibility of finding isolated critical points and the associated nonstandard critical exponents. As we will see, we encounter unexpected cases, which are absent in Lovelock gravity theories. Finally, in Sec. \ref{sect4:conclusion}, we summarize the results and finish our paper with some concluding remarks.

\section{$4D$ Lovelock field equations, $\text{AdS}_4$ black holes, and extended phase space thermodynamics} \label{sect2:Regularized Lovelock}

The Lagrangian of Lovelock gravity is
 given by a sum of dimensionally extended Euler densities as \cite{Lovelock1971,Lovelock1972}
 \begin{equation} \label{Lagrangian}
 {I_{{\rm{Lovelock}}}} = \int {{d^D}x\sqrt { - g} \sum\limits_{m = 0}^N {{\alpha _m}{{\cal L}_m}} },
 \end{equation}
 \begin{equation}
 {{\cal L}_m} = \frac{1}{{{2^m}}}\delta _{{\rho _1}\,{\sigma _1}\,...\,{\rho _m}\,{\sigma _m}}^{{\mu _1}\,{\nu _1}\,...\,{\mu _m}\,{\nu _m}}{R_{{\mu _1}\,{\nu _1}}}^{{\rho _1}\,{\sigma _1}}...\,{R_{{\mu _m}\,{\nu _m}}}^{{\rho _m}\,{\sigma _m}}
 \end{equation}
where $\delta _{{\rho _1}\,{\sigma _1}\,...\,{\rho _p}\,{\sigma
 		_p}}^{{\mu _1}\,{\nu _1}\,...\,{\mu _p}\,{\nu _p}}$ and ${R_{{\mu_p}\,{\nu _k}}}^{{\rho _p}\,{\sigma _p}}$ are the generalized
 totally antisymmetric Kronecker delta and the Riemann tensor respectively. The upper bound on the sum, i.e., $N = \left[ {\frac{{D - 1}}{2}} \right]$, specifies the maximal order of higher-curvature corrections in a $D$-dimensional spacetime, in which the closed brackets $[...]$ denote taking the integer part. The first term in the Lagrangian (\ref{Lagrangian}), $\alpha_0 {\cal L}_0$, gives the cosmological constant and the second term is proportional to Ricci scalar, $\alpha_1 {\cal L}_1=R$ (we have set $\alpha_1 =1$ in order to recover Einstein gravity in the absence of higher-order curvature corrections.) The correction terms such as ${\cal L}_m$'s with $2m > D$ vanishes identically and those terms with $2m = D$ are purely topological and do not contribute to the gravitational field equations. So, in a given dimension, there exist always a finite number of dimensionful parameters, $\alpha_m$'s. The $4D$ regularization of Lovelock gravity can be obtained by use of rescaling the Lovelock couplings ($\alpha_m$'s) as \cite{Glavan2020,Zhidenko2020a,Casalino2021}
 \begin{equation} \label{rescaling}
 {\alpha _m} \to {\alpha _m}\prod\limits_{j = 3}^{2m} {\frac{1}{{D - j}}},
 \end{equation}
 which leads to higher-order curvature terms also contributing to the four-dimensional gravitational field equations. Assuming the static ansatz for the black hole spacetime with horizon being a surface of constant scalar curvature ($k=+1$ spherical, $k=0$ Ricci flat, and $k=-1$ hyperbolic topologies) as
 \begin{equation} \label{ansatz}
 d{s^2} =  - f(r)d{t^2} + \frac{{d{r^2}}}{{f(r)}} + {r^2}\Bigg( {dx_1^2 + \frac{{{{\sin }^2}(\sqrt k {x_1})}}{k}\sum\limits_{i = 2}^{D - 2} {dx_i^2\prod\limits_{j = 2}^{i - 1} {{{\sin }^2}{x_j}} } } \Bigg),
 \end{equation}
 along with employing the rescaling (\ref{rescaling}) and then taking the $D \to 4$ limit, one obtains the reduced gravitational field equation, given by \cite{Zhidenko2020a,Casalino2021}
 \begin{equation} \label{reduced FE}
\sum\limits_{m = 1}^N {{\alpha _m}{{\left( {\frac{{k - f(r)}}{{{r^2}}}} \right)}^m}}  = \frac{\Lambda }{3} + \frac{M}{{{r^3}}},
 \end{equation}
where $M$ is an integration constant related to the black hole's ADM mass. When the Lovelock couplings ($\alpha_m$'s with $m \ge 2$) obey the following special relationships \cite{ICP2014}
   \begin{equation} \label{special couplings}
 {\alpha _m} = {\alpha _N}{\left( {\frac{1}{{N{\alpha _N}}}} \right)^{\frac{{N - m}}{{N - 1}}}}\left( {\begin{array}{*{20}{c}}
 	N\\
 	m
 	\end{array}} \right),
 \end{equation}
the reduced gravitational field equation in the conventional Lovelock gravity can be analytically resolved, which leads to an unusual Taylor expansion for the equation of state of hyperbolic black holes, which eventually results in nonstandard critical exponents associated with the isolated critical points (i.e., the inflection points of of the finely tuned hyperbolic black holes) \cite{ICP2014,Frassino2014Mann}. We confirmed that such fine-tunings are also essential to have such an unusual Taylor expansion in the regularized $4D$ theory. Assuming the fine-tunings (\ref{special couplings}), we confirmed that the reduced gravitational field equation (\ref{reduced FE}) can be analytically resolved, yielding
 \begin{equation} \label{metric function}
 f(r) = k + \frac{{{r^2}}}{{{{\left( {N{\alpha _N}} \right)}^{\frac{1}{{N - 1}}}}}}\left[ {1 - {{\left( {1 + {{\left( {{N^N}{\alpha _N}} \right)}^{\frac{1}{{N - 1}}}}\left( {\frac{M}{{{r^3}}} + \frac{\Lambda }{3}} \right)} \right)}^{\frac{1}{N}}}} \right].
 \end{equation}
The above reduced field equation is exactly the $D \to 4$ limit of its counterpart in Lovelock gravity (eq. (9) in Ref. \cite{ICP2014}) and note that it is valid for any order $N$. Assuming the parameters are properly selected, the largest root of the metric function (\ref{metric function}) is the location of the event (outer) horizon, which we denote by $r=r_+$. The AdS asymptotic behavior ensures a positive slope for ${\left. {\left( {\partial f(r)/\partial r} \right)} \right|_{r = {r_ + }}}$, leading to a positive Hawking temperature. Details of the finely tuned $\text{AdS}_4$ black hole solutions (\ref{metric function}) can be found in Ref. \cite{PV4DLovelock}. In general, for $\text{AdS}_4$ black holes (not only the finely tuned solutions), i.e., the $D \to 4$ limit of ansatz (\ref{ansatz}) in the regularized $4D$ Einstein-Lovelock gravity, the thermodynamic quantities including the ADM mass ($M$), the entropy ($S$), and the temperature ($T$) read \cite{Casalino2021,Zhidenko2020a,PV4DLovelock}
\begin{equation} \label{mass}
{\cal M} = \frac{{{V_k}}}{{8\pi }}M = \frac{{{V_k}}}{{8\pi }}\left[ {\sum\limits_{m = 1}^N {{\alpha _m}{k^m}r_ + ^{3 - 2m} - \frac{\Lambda }{3}r_ + ^3} } \right],
\end{equation}
 \begin{equation} \label{entropy}
S = \frac{{{V_k}}}{4}\sum\limits_{m = 1\,,\, \ne 2}^N {{\alpha _m}\left( {\frac{{m{k^{m - 1}}}}{{2 - m}}} \right)r_ + ^{4 - 2m} + {V_k}k{\alpha _2}\ln \left( {\frac{{{r_ + }}}{{\sqrt {{\alpha _2}} }}} \right)},
 \end{equation}
  \begin{equation} \label{temp}
 T = \frac{{\sum\limits_{m = 1}^N {k{\alpha _m}(3 - 2m){{\left( {\frac{k}{{r_ + ^2}}} \right)}^{m - 1}} - \Lambda r_ + ^2} }}{{4\pi {r_ + }\left( {\sum\limits_{m = 1}^N {m{\alpha _m}{{\left( {\frac{k}{{r_ + ^2}}} \right)}^{m - 1}}} } \right)}}.
 \end{equation}
Interpreting the negative cosmological constant ($\Lambda$) as a positive thermodynamic pressure ($P$) leads to  extending the thermodynamic phase space of black holes \cite{Kastor2009,Kastor2010,KubiznakMann2012,CQG2017Review}. By this extension, known as the extended phase space thermodynamics, the pressure and subsequently the thermodynamic volume are defined as
 \begin{equation}
 P =  - \frac{\Lambda }{{8\pi }}, \qquad V = \left( {\frac{{\partial {\cal M}}}{{\partial P}}} \right) = \frac{{{V_k}}}{3}r_ + ^3.
 \end{equation}
Naturally, the ADM mass of AdS black holes is identified as the chemical enthalpy of the black hole system. The first law of extended black hole thermodynamics and the corresponding Smarr formula are obtained as
\begin{equation}
d{\cal M} = TdS - VdP + \sum\limits_{m = 2}^N {{\Psi _m}d} {\alpha _m},
\end{equation}
 and
\begin{equation}
 {\cal M} = 2TS - 2PV + \sum\limits_{m = 2}^N {2(m - 1)} {\Psi _m}{\alpha _m},
\end{equation}
 where the potentials ${\Psi _m}$ are the thermodynamic conjugates to the Lovelock couplings ($\alpha_m$), given by
  \begin{equation}
 {\Psi _m} = {\left( {\frac{{\partial M}}{{\partial {\alpha _m}}}} \right)_{S,P,Q,{\alpha _{n \ne m}}}}.
 \end{equation}
The thermodynamic description for the fine-tuned case of Lovelock gravity theories that we are dealing with is simply obtained by applying eq. (\ref{special couplings}) for the Lovelock couplings.

\section{Looking for isolated critical points with nonstandard critical exponents} \label{sect3:Absence}

In order to study the critical behavior of the black hole system, we first obtain the corresponding equation of state by inserting $\Lambda = -8 \pi P$ into the Hawking temperature (\ref{temp}) and rearranging for pressure, yielding 
 \begin{equation} \label{pressure}
 P = \frac{1}{{8\pi }}\sum\limits_{m = 1}^N {\frac{{{\alpha _m}}}{{r_ + ^2}}{{\left( {\frac{k}{{r_ + ^2}}} \right)}^{m - 1}}\left[ {4\pi m{r_ + }T - k(3 - 2m)} \right]}.
 \end{equation}
The critical points (if exist any) are determined by finding the inflection points of the isothermal curves in $P-V$ diagrams and can be found from the following relations
 \begin{equation} \label{critical Eq}
 {\left( {\frac{{\partial P}}{{\partial v}}} \right)_{T,\,...}} = {\left( {\frac{{{\partial ^2}P}}{{\partial {v^2}}}} \right)_{T,\,...}} = 0\,\,\, \leftrightarrow \,\,\,{\left( {\frac{{\partial P}}{{\partial {r_ + }}}} \right)_{T,\,.\,..}} = {\left( {\frac{{{\partial ^2}P}}{{\partial {r_ + }^2}}} \right)_{T,\,...}} = 0,
 \end{equation}
 where $v$ is the specific volume proportional to $r_+$ as $v=2r_+ \ell_\text{P}^2$ \cite{KubiznakMann2012,CQG2017Review,Mann2012Altamirano,Frassino2014Mann}. ($\ell_\text{P}$ is the Planck length and hereafter is set to one.) In Lovelock gravity theories of odd order $N$ and $k=-1$ (hyperbolic symmetry), the second-order critical points which satisfy eq. (\ref{critical Eq}) are called isolated critical points.\footnote{The authors in Refs. \cite{HennigarTjoa2017,Dykaar2017JHEP,Mir2019b}, by extending Lovelock gravity, have constructed a broader class of quasi-topological gravity theories with(out) conformal scalar hair, in which isolated critical points can be found for spherical black holes. But this does not take place in conventional Lovelock gravity and also the regularized $4D$ theory under consideration.} In order to check whether there is a possibility of finding isolated critical points in the presence of higher-order curvatures, eq. \ref{critical Eq} can analytically be solved for arbitrary (odd) $N$. Here we can use the same technical term but this, for the reasons that will discussed later, may lead to misunderstandings. So, we prefer to refer to them as \textit{isolated special points} or simply \textit{inflection points}. The thermodynamic data associated with the isolated special points of the hyperbolic vacuum black holes in the $N$th (odd) order of $4D$ Einstein-Lovelock gravity are obtained as
 \begin{equation} \label{isolated SP-N}
 {r_i} = {\left( {N{\alpha _N}} \right)^{\frac{1}{{2(N - 1)}}}}\,,\,\,\,\,{T_i} = \frac{1}{{2\pi {r_i}}}\,,\,\,\,\,{P_i} = \frac{{3{\alpha _N}}}{{8\pi }}{\left( {\frac{1}{{N{\alpha _N}}}} \right)^{\frac{N}{{N - 1}}}}.
 \end{equation}
Eq. (\ref{isolated SP-N}) is exactly the $D \to 4$ limit of its counterpart in Lovelock gravity (see Ref. \cite{ICP2014}). At this special point, the black hole mass vanishes for any odd order $N$. The obtained pressure at this special point, $P_i$, in fact is the maximum allowable pressure, which can be confirmed by studying the asymptotic behavior of the metric function (\ref{metric function}),
\begin{equation}
{\left. {f(r)} \right|_{{\rm{large}}\,r}} = k + \frac{{{r^2}}}{{{{\left( {N{\alpha _N}} \right)}^{\frac{1}{{N - 1}}}}}}\left[ {1 - {{\left( {1 - \frac{{8\pi P}}{3}{{\left( {{N^N}{\alpha _N}} \right)}^{\frac{1}{{N - 1}}}}} \right)}^{\frac{1}{N}}}} \right]\,\,\,\, \to \,\,\,\,{P_{\max }} = {P_i}\,.
\end{equation}
For pressures above this bound, i.e., $P > P_i$, the black hole spacetime is no longer asymptotically AdS and the metric function (\ref{metric function}) becomes imaginary. Expanding the equation of states in the vicinity of isolated special point yields \footnote{As usual, we have utilized the following reduced variables
	\begin{equation}
	p = \frac{P}{{{P_i}}} \, , \quad
	\tau  \equiv \frac{{T - {T_i}}}{{{T_i}}} \, , \quad  w \equiv \frac{{v - {v_i}}}{{{v_i}}} = \frac{{{r_ + } - {r_i}}}{{{r_i}}}.
	\end{equation}}
\begin{equation}
p \equiv \frac{P}{{{P_i}}} = 1 + \frac{{{2^N}N}}{3}\tau {w^{N - 1}} + \frac{{{2^N}(N - 3)}}{3}{w^N} + ...,
\end{equation}
at first glance, suggesting a set of nonstandard critical exponents the same as those in hyperbolic vacuum black holes of Lovelock gravity. But, we confirmed that, for any $N$($> 3$)th odd-order Lovelock vacuum black holes with hyperbolic geometry ($k=-1$), the corresponding entropy at the inflection point (\ref{isolated SP-N}) is negative, meaning that all of these special points (for $N>3$) are nonphysical. Note that there exist ($N-2$) inflection points for eq. (\ref{critical Eq}) and the point associated with the smaller horizon's radius is the counterpart of the isolated critical points in $N$th odd-order Lovelock gravity. The entropy associated with the another inflection points (for any $N > 3$) is negative again, so there is no criticality associated with these points as well. \vspace{2mm}

The story is quite different for the case of hyperbolic vacuum black holes in the regularized $4D$ Einstein-Lovelock gravity with cubic curvature corrections ($N=3$). In this case, there is only one inflection point, given by
 \begin{equation} \label{isolated CP-3}
 {r_i} = \sqrt \alpha  ,\,\,{T_i} = \frac{1}{{2\pi \sqrt \alpha  }},\,\,\,{P_i} = \frac{1}{{8\pi \alpha }}.
 \end{equation}
 In the vicinity of this point, the physical properties of the black hole system change drastically. Interestingly, at this point, both the mass and the entropy of the black hole system vanishes and one obtains the following Taylor expansion for the equation of state 
 \begin{equation} \label{third order EoS}
 p = 1 + 8\tau {\omega ^2} - 32\tau {\omega ^3} + (82\tau  - 4){\omega ^4} + (20 - 170\tau ){\omega ^5} + O\left( {\tau {\omega ^6},{\omega ^6}} \right),
 \end{equation}
 which clearly would lead to a new set of critical exponents if there is criticality (a phase transition) associated with the isolated special point (\ref{isolated CP-3}).\footnote{ For example, the independent critical exponent $\delta$ is related to the behavior of the critical isotherm $T=T_c$ in $P-v$ diagram, through
 \begin{equation}
	\left| {P - {P_c}} \right| \propto {\left| {v - {v_c}} \right|^\delta }.
 \end{equation}
 Putting $\tau = 0$ (i.e., $T=T_i$) in eq. (\ref{third order EoS}), it follows that we have $\delta = 4$. On the other hand, the critical exponent, $\beta$, determines the behavior of the order parameter $\eta$, given by
 	\begin{equation}
 	\eta  = {v_ > } - {v_ < } \propto {\left| \tau  \right|^\beta },
 	\end{equation}
 in which $v_ >$ and $v_ < $ are the volumes of the black hole in large and small phase states, respectively. The equation of state (\ref{third order EoS}) implies that $\beta = 1/2$. These two independent exponents are enough to ensure a violation of the scaling relations and having a completely new set of nonstandard exponents. But we argue that there is no criticality associated with this special point.} \vspace{2mm}

\begin{figure}[!htbp]
	$%
	\begin{array}{ccc}
	\epsfxsize=5.6cm \epsffile{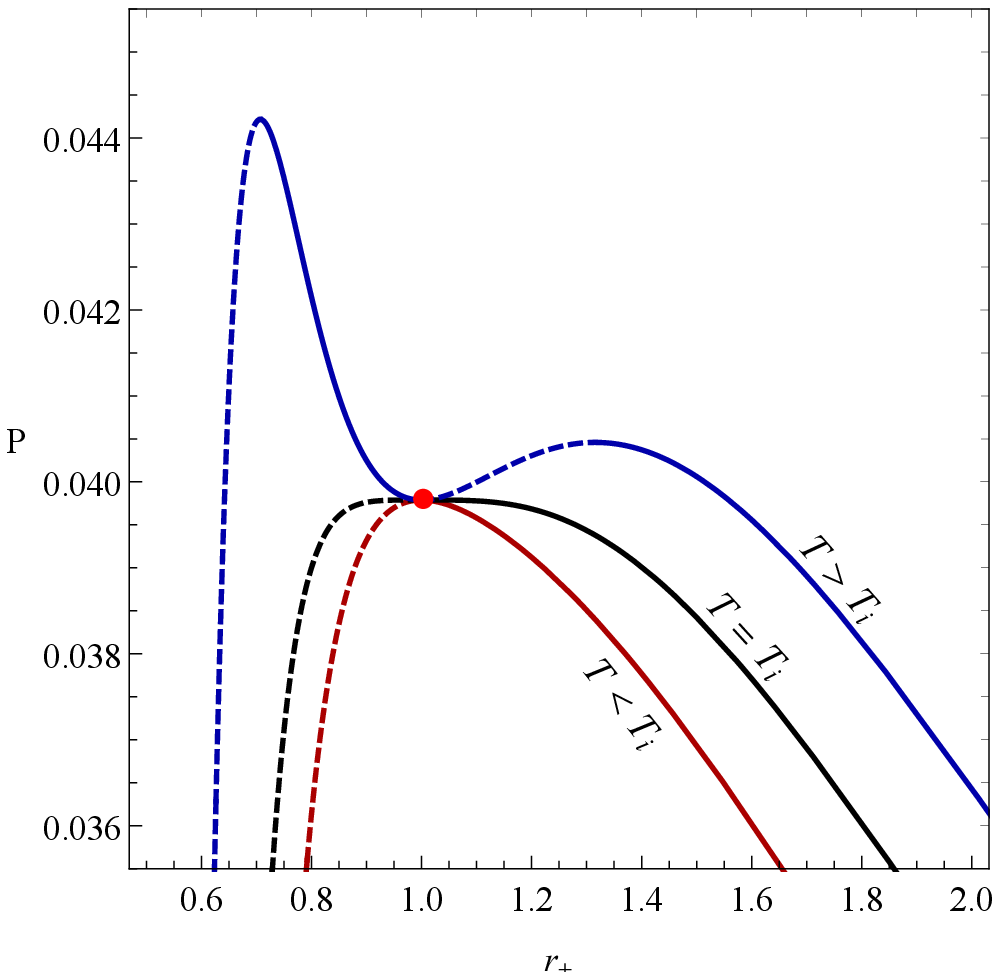}
	\epsfxsize=5.6cm \epsffile{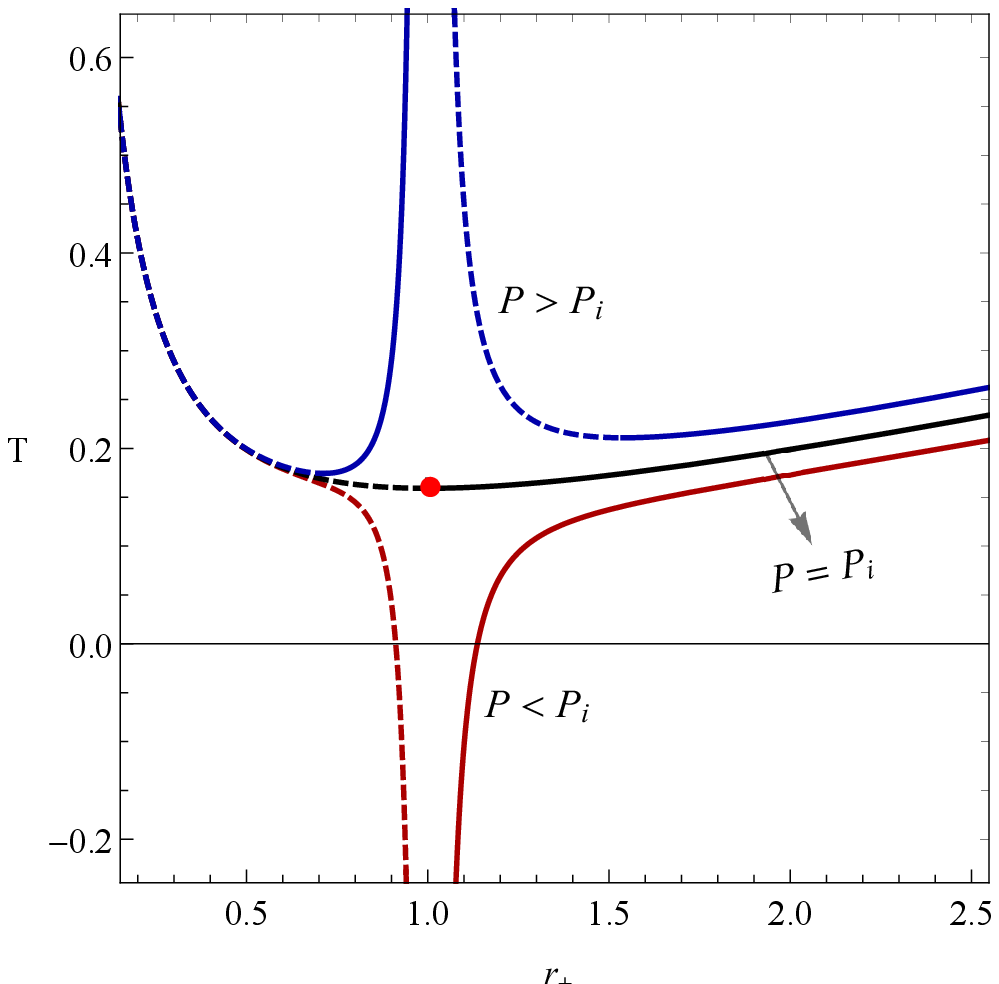}
	\epsfxsize=5.9cm \epsffile{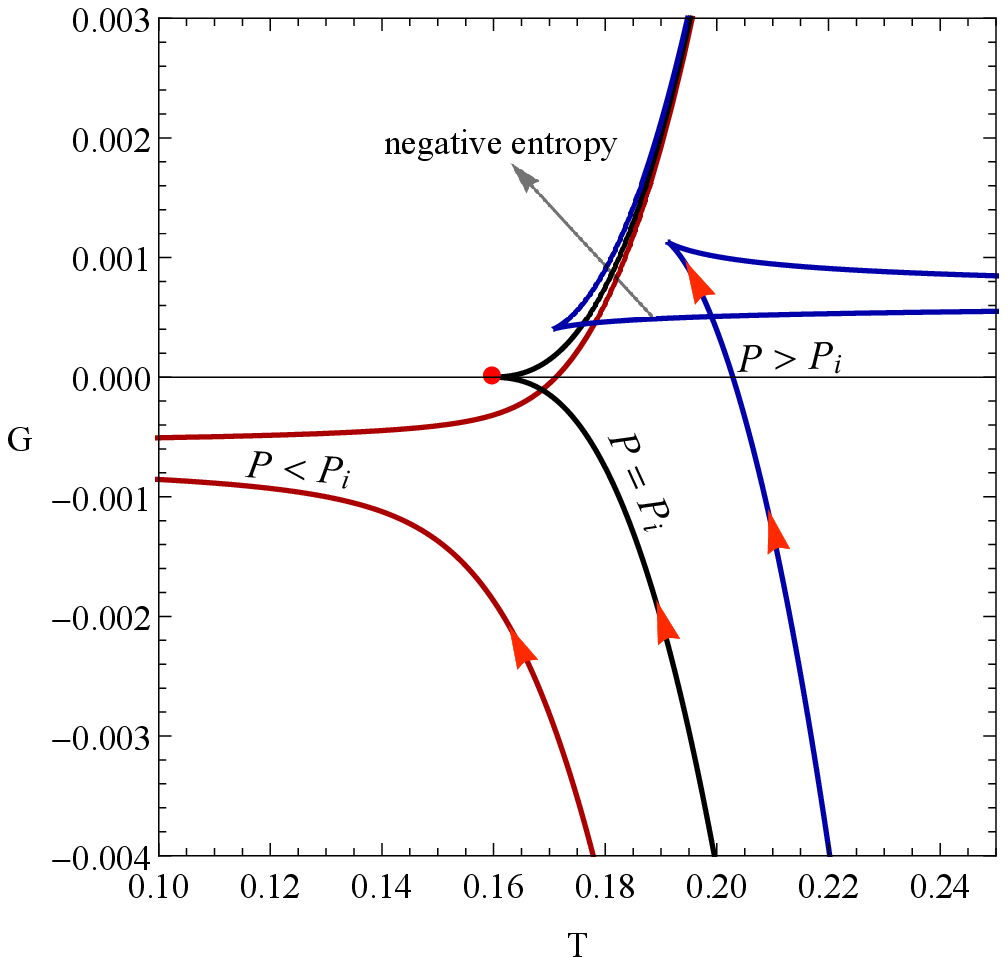}
	&  &
	\end{array}
	$%
	\caption{\textbf{The behavior of phase diagrams near the isolated special point ($r_i$, $T_i$, $P_i$) in hyperbolically symmetric (vacuum) black holes:} $P-r_{+}$ (left), $T-r_{+}$
		(middle) and $G-T$ (right) diagrams; we have set $k=-1$, $\alpha=1$, and $N=3$. The red circles correspond to the isolated special point (the inflection point in $P-r_+$ diagram). Dashed lines in the $P-r_+$ diagram represents negative isothermal compressibility, $\kappa_T<0$, which is not physical. Dashed lines in the $T-r_+$ diagram correspond to negative isobaric expansivity, $\beta_P<0$, or equivalently negative heat capacity ($C_P<0$). \textit{Thermodynamic data}: ($T_i \approx 0.1592$, $P_i \approx 0.03979$, and $r_i=1$).}
	\label{PV-Hyper-special}
\end{figure}

Further insights about this issue can be gained by analyzing the phase diagrams near this special point. To do so, we have depicted $P-r_+$, $T-r_+$, and $G-T$ diagrams in Fig. \ref{PV-Hyper-special}. Let us first focus on the $P-r_+$ diagram. The dashed curves represent the nonphysical parts of isotherms since the isothermal compressibility, i.e.,
${\kappa _T} =  - \frac{1}{v}{\left( {\frac{{\partial v}}{{\partial P}}} \right)_T}$, is negative. It is not possible that the (unphysical) oscillatory part of isotherms with $T > T_i$ can be replaced by a line of constant pressure according to the Maxwell’s equal-area law because any attempt in this direction will cause the black hole's pressure to exceed the maximal pressure. For $T \le T_i$, the isotherms for $r \ge r_i$ are physical and have strictly decreasing behaviors. The location of merging isotherms in the $P-r_+$ diagram also determines the so-called thermodynamic singularity \cite{Frassino2014Mann} and it can be found using the following condition
\begin{equation} \label{thermodynamic singularity}
{\left. {\frac{{\partial P}}{{\partial T}}} \right|_{{r_ + } = {r_s}}} = 0\,\,\,\, \to \,\,\,\,\sum\limits_{m = 1}^N {m{\alpha _m}{{\left( {\frac{k}{{r_ s ^2}}} \right)}^{m - 1}}}  = 0.
\end{equation}
We confirmed that in any regularized $4D$ Einstein-Lovelock theory of odd order $N$, the isolated special point (\ref{isolated SP-N}) is in fact a thermodynamic singular point, satisfying Eq. (\ref{thermodynamic singularity}). \vspace{2mm}

Regarding the $T-r_+$ diagram in Fig. \ref{PV-Hyper-special}, the isobars which are above the maximal pressure bound ($P > P_i$), are not allowed. Dashed lines in the $T-r_+$ diagram correspond to negative isobaric expansivity, $\beta  = \frac{1}{v}{\left( {\frac{{\partial v}}{{\partial T}}} \right)_P} < 0$, or equivalently negative specific heat at constant pressure ($C_P < 0$). For the isobar with $P = P_i$, despite the nonphysical part, a strictly increasing behavior is observed. For isobaric curves in the range $P \le P_i$, there exist always a lower bound for the radius of the event horizon for which the temperature is always positive and the isobaric expansivity takes physical values (with positive sign). At $r_+=r_s$, the temperature diverges for all the isobars except for the $P=P_i$ isobar. \vspace{2mm}

Considering the $G-T$ diagram in Fig. \ref{PV-Hyper-special}, it is observed that, for isobaric curves in the range $P < P_i$, the global minimum of the Gibbs free energy has a strictly decreasing behavior as the temperature increases, which rules out possibility of any phase transition. The isotherms with pressures in the range $P > P_i$ exceed the maximal pressure (\ref{isolated CP-3}), so they are nonphysical again. For the $P = P_i$ isobar, neither the first derivative the Gibbs free energy ($G$) nor its second derivative exhibit any jump or discontinuity, meaning that there is no criticality associated with the inflection point (\ref{isolated CP-3}).  In cases with $N > 3$, qualitatively, the same behavior is observed, but the inflection point occurs at the branch with negative entropy (the branch with negative slope in the corresponding $G-T$ diagram). By increasing the temperature, the black hole entropy eventually becomes physical (positive) since the global minimum of the Gibbs energy in this part has a strictly decreasing behavior, implying $S=-\frac{\partial G}{\partial T} >0$. These are in strong contrast to Lovelock gravity. As shown in Ref. \cite{ICP2014}, in the standard Lovelock gravity, the Gibbs free energy displays two swallowtails, both emerging from the isolated critical point. \vspace{2mm}

\begin{figure}[!htbp]
	$%
	\epsfxsize=8cm \epsffile{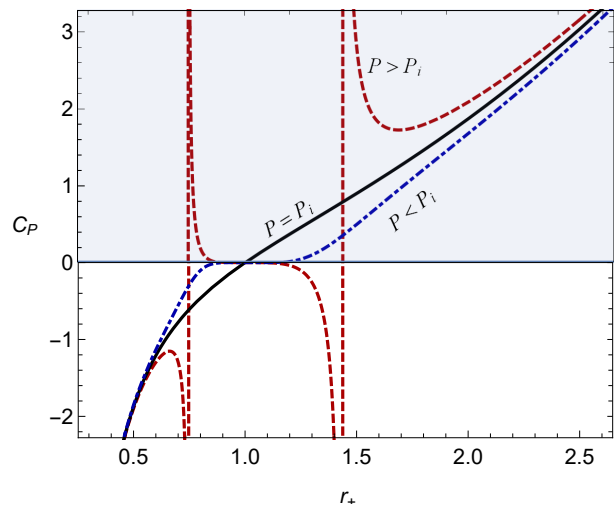}
	$%
	\caption{The behavior of specific heat , $C_P$ versus $r_+$, near the isolated special point ($r_i$, $T_i$, $P_i$) in hyperbolically symmetric (vacuum) black holes. We have set $k=-1$, $\alpha=1$, and $N=3$.}
	\label{HC}
\end{figure}

  Therefore, there exist several reasons preventing us to interpret the special point (\ref{isolated CP-3}) as an isolated (second-order) critical point with nonstandard critical exponents:

\begin{itemize}
	
\item i) The first-order coexistence lines are terminated at the second-order critical points. By analyzing the phase diagrams in Fig. \ref{PV-Hyper-special}, we showed that the inflection point (\ref{isolated CP-3}) is not the endpoint of any first-order coexistence line and neither upcritical isobars nor subcritical isobars exhibit any phase transition. This lead us to conclude that there is no criticality associated with this point.

\item ii) Since the entropy of the black hole system vanishes at the isolated special point, there do not exist degrees of freedom in order for a phase transition to occur. This means there is no criticality and phase transition associated with this special point, as supported by the concepts of statistical mechanics. 

\item iii) Second-order critical points lead to discontinuities in some thermodynamical quantities such as specific heat, therefore a divergence point of specific is a sign of a possible phase transition. This is due to that the critical point occurs at the spike-like divergence of the specific heat at constant pressure, i.e., they are inflection points in $P-V$ diagrams. In Fig. \ref{HC}, the behavior of specific heat at constant pressure versus the horizon's radius (or equivalently specific volume $v$) is displayed for different pressures. As seen, for $P = P_i$, the specific heat is an increasing function of specific volume $v$ ($\propto r_+$) and at $r=r_i$ the specific heat vanishes.
\end{itemize}

\section{Closing remarks} \label{sect4:conclusion}

We presented the finely tuned $4D$ Einstein-Lovelock black hole solutions which include an arbitrary number of higher-curvature corrections. Only in this case, the equation of state of $\text{AdS}_4$ black holes with hyperbolic symmetry ($k=-1$) can take a special form, yielding an unusual Taylor expansion about the possible (isolated) critical point. We showed that the the equation of state of hyperbolic vacuum black holes in the regularized $4D$ Einstein-Lovelock gravity theories of odd order $N > 3$ do not possess any physical inflection point since the corresponding entropy is always negative definite. It is also proved that, by taking the $D \to 4$ limit of Lovelock gravity of odd order $N =3 $, the isolated critical points and the special property of having nonstandard critical exponents in the theory disappear (the reasons have been listed in Sect. \ref{sect3:Absence}). In conclusion, isolated critical points with nonstandard critical exponents do not exist in the four-dimensional regularization of Lovelock gravity. The obtained results are nontrivial since all the known examples of nonstandard critical exponents in black hole physics are relied on the correction terms arisen from Lovelock Lagrangian of odd order in curvature \cite{ICP2014,Frassino2014Mann,HennigarTjoa2017,Dykaar2017JHEP,Mir2019a,Mir2019b,Brenna2015}. The results so far found in literature have been summarized in Table \ref{tab:summary} for the readers' convenience. \vspace{2mm}

\begin{table}[]
	\caption{Summary of standard and nonstandard critical exponents in vdW fluid and different black hole systems.}
	\label{tab:summary}
	\begin{tabular}{|c|c|c|c|c|}
		\hline
		Critical exponents                                                                                                                                            & $\alpha$ & $\beta$ & $\gamma$ & $\delta$ \\ \hline
		Van der Waals fluid \cite{HuangBook}                                                                                                                                   & 0     & 1/2  & 1     & 3     \\ \hline
		\begin{tabular}[c]{@{}c@{}}
			RN-AdS, Kerr-AdS, ... black holes \cite{KubiznakMann2012,CQG2017Review,Mann2012Altamirano,HennigarMann2017PRL,CQG2020,DH2021,Hennigar2017JHEP,Zou2014,Galaxies2014}and also\\
			spherical black holes in Lovelock gravity and cubic quasi-topological\\
			gravity theories with(out) conformal scalar hair \cite{Frassino2014Mann,HennigarTjoa2017,Dykaar2017JHEP,Mir2019a,Mir2019b,Brenna2015}                        \end{tabular}                                                                                           & 0     & 1/2  & 1     & 3     \\ \hline 
		\begin{tabular}[c]{@{}c@{}}Hyperbolic (vacuum) black holes\\ in Lovelock gravity theories of odd order $N$ \cite{Frassino2014Mann,ICP2014} \end{tabular}                                         & 0     & 1    & $N-1$   & $N$     \\ \hline
		\begin{tabular}[c]{@{}c@{}}Hyperbolic (vacuum) black holes in the 4D regularization\\ of Lovelock gravity theories of odd order $N$ (the present study) \end{tabular}       & -     & -    & -     & -     \\ \hline
		\begin{tabular}[c]{@{}c@{}}Hyperbolic (vacuum) AdS black holes\\ in cubic ($N=3$) quasi-topological gravity \cite{Brenna2015} \end{tabular}                                 & 0     & 1    & 2     & 3     \\ \hline
		\begin{tabular}[c]{@{}c@{}}Hyperbolic and spherical AdS black holes in\\ cubic ($N=3$) Lovelock gravity with conformal scalar hair \cite{HennigarTjoa2017} \end{tabular}     & 0     & 1    & 2     & 3     \\ \hline
		\begin{tabular}[c]{@{}c@{}}Hyperbolic and spherical AdS black holes in cubic\\($N=3$) quasi-topological gravity with conformal scalar hair \cite{Dykaar2017JHEP} \end{tabular} & 0     & 1    & 2     & 3     \\ \hline
		\begin{tabular}[c]{@{}c@{}}Hyperbolic and spherical (vacuum) AdS black holes in\\ cubic ($N=3$) generalized quasi-topological gravity \cite{Mir2019a,Mir2019b} \end{tabular}                    & 0     & 1    & 2     & 3     \\ \hline
	\end{tabular}
\end{table}

In all other cases, assuming any arbitrary number of higher-curvature corrections, we found that the equation of state around the critical point has the following Taylor expansion
\begin{equation}
p \equiv \frac{P}{{{P_c}}} = 1 + A \tau  +B \tau w + C w^3 + ...,
\end{equation}
which is similar to those of the conventional Lovelock gravity \cite{Frassino2014Mann,Zou2014}, massive gravity \cite{CQG2020,DH2021} etc. \cite{Hennigar2017JHEP,Galaxies2014}. This is a well-known Taylor expansion which always gives the mean-field critical exponents as $\alpha = 0$, $\beta = 1/2$, $\gamma = 1$, and $\delta = 3$. Interestingly, the final results do not depend on the functions $A$, $B$, $C$ etc., so the details of them are not important at all. Note that, these functions take different forms in different theories as well as in different ensembles \cite{CQG2020,DH2021,Zou2014}. \vspace{2mm}

The four-dimensional regularization procedure that we were dealing with in this paper was a method recently employed by Glavan and Lin \cite{Glavan2020} (also see the earlier works of Refs. \cite{Tomozawa2011,Zerbini2013}). As mentioned in Sect. \ref{sect1:intro}, some subtleties and criticisms on the $D \to 4$ limit of Gauss-Bonnet gravity (and naturally the $D \to 4$ limit of Lovelock gravity), upon rescaling the Lovelock coupling constants $\alpha$'s according to eq. (\ref{rescaling}), were revealed by several authors \cite{Tekin2020,Mann2020EGB4D,Mahapatra,Ai2020,Hinterbichler2020,Kobayashi2020,LuPang2020,Mukohyama2020,EGB4Dreview} after the proposal of Ref. \cite{Glavan2020}. In this regard, alternative versions of regularized $4D$ Lovelock gravity theories have been constructed using conformal regularization \cite{Mann2020EGB4D}, regularized Kaluza-Klein reduction \cite{Kobayashi2020,LuPang2020}, and temporal diffeomorphism breaking regularization \cite{Mukohyama2020} etc. \cite{EGB4Dreview}. As a result, it is possible that the results of this research may depend on the regularization procedure that we selected and this could be the reason the isolated critical points fail to exist. As long as black hole solutions and the corresponding thermodynamic quantities from the alternative regularized $4D$ Lovelock gravity theories are similar to those of the naive $D \to 4$ limit of $D$-dimensional Lovelock gravity upon rescaling the Lovelock couplings (\ref{rescaling}), trivially the result of this research is valid for them. In this regard, it should be emphasized that the $4D$ limit of Gauss-Bonnet gravity by means of conformal regularization as well as regularized Kaluza-Klein reduction share solutions with the Glavan and Lin's formulation of $4D$ Einstein-Gauss-Bonnet gravity \cite{Mann2020EGB4D,LuPang2020,EGB4Dreview}. Furthermore, up to cubic curvature corrections ($N=3$), the $D \to 4$ limit of cubic Lovelock gravity through a regularized Kaluza-Klein reduction revealed that black hole solutions with planar horizon similar to those of the naive $4D$ theory (based on the Glavan and Lin's proposal, Sect. \ref{sect2:Regularized Lovelock}) are allowed \cite{Alkac2022} which obviously rule out an existence of isolated critical point again. (Whether or not black hole solutions with spherical or hyperbolic horizon exist for this regularized theory remains an open question. An existence of isolated critical points crucially depends on such solutions.) In order to examine the possibility of any isolated critical point with nonstandard critical exponents, we need to construct topological black hole solutions (of odd order $N$) with higher-order corrections of spacetime curvature beyond Gauss-Bonnet, as presented by us in this work, and so far no such solutions have been presented for other regularization methods. \vspace{2mm}

Other nontrivial results are expected for the $4D$ regularization of Gauss-Bonnet and Lovelock gravity theories, e.g., one can disclose that taking the $D \to 4$ limit of Gauss-Bonnet gravity (Lovelock gravity up to quadratic curvature corrections, $N=2$) leads to having criticality and novel phase transitions for hyperbolic vacuum black holes. This is another nontrivial result of taking the $D \to 4$ limit of Lovelock gravity theories since hyperbolic vacuum (or charged) black holes in Gauss-Bonnet gravity do not possess any critical behavior.  Further investigations show that, the so-called thermodynamic singularity, which was unphysical in Gauss-Bonnet gravity \cite{Frassino2014Mann}, becomes physical in the $D \to 4$ limit of Gauss-Bonnet gravity. The detailed study of this case is postponed for future study. These results are highly nontrivial and motivate us to further explore the subject of extended black hole thermodynamics in the $D \to 4$ limit of Lovelock gravity theories in future. We are currently investigating this issue in Ref. \cite{PV4DLovelock}. Furthermore, it becomes of interest to investigate the possibility of the existence of nonstandard critical exponents in BTZ black holes with higher curvature corrections through the $3D$ regularization of Lovelock gravity. For this purpose, the regularized BTZ Lovelock black hole solutions can be found in Ref. \cite{Zhidenko3DBTZ}, where they were constructed and studied in detail.

\begin{acknowledgements}
We gratefully thank the anonymous referee for enlightening comments and suggestions which substantially helped in improving the quality of the paper. Sadly, after this work was completed, a tragic event led us to mourn the loss of Prof. M.R. Setare. We would like to express our deepest condolences on his passing. A.D. wish to thank the support of Iran Science Elites Federation (ISEF).\\

\textbf{Data Availability Statement} Data sharing is not applicable, as no data sets were generated or analyzed in this research.
\end{acknowledgements}

\end{document}